\documentstyle{BSAXwork}

\input epsf.sty

\def \AAP #1 #2 {{\em Astron. Astrophys.\/} {\bf #1}, #2}
\def \AAL #1 #2 {{\em Astron. Astrophys. Lett.\/} {\bf #1}, L#2}
\def \AAR #1 #2 {{\em Astron. Astrophys. Rev.\/} {\bf #1}, #2}
\def \AAS #1 #2 {{\em Astron. Astrophys. Suppl. Ser.\/} {\bf #1}, #2}
\def \AJ #1 #2 {{\em Astron. J.\/} {\bf #1}, #2}
\def \ANNREV #1 #2 {{\em Ann. Rev. Astron. Astrophys.\/} {\bf #1}, #2}
\def \APJ #1 #2 {{\em Astrophys. J.\/} {\bf #1}, #2}
\def \APJL #1 #2 {{\em Astrophys. J. Lett.\/} {\bf #1}, L#2}
\def \APJS #1 #2 {{\em Astrophys. J. Suppl.\/} {\bf #1}, #2}
\def \APSS #1 #2 {{\em Astrophys. Space Sci.\/} {\bf #1}, #2}
\def \ASR #1 #2 {{\em Adv. Space Res.\/} {\bf #1}, #2}
\def \BAIC #1 #2 {{\em Bull. Astron. Inst. Czechosl.\/} {\bf #1}, #2}
\def \JSQRT #1 #2 {{\em J. Quant. Spectrosc. Radiat. Transfer\/} {\bf #1}, #2}
\def \MN #1 #2 {{\em Mon. Not. R. Astr. Soc.\/} {\bf #1}, #2}
\def \MEM #1 #2 {{\em Mem. R. Astr. Soc.\/} {\bf #1}, #2}
\def \PLR #1 #2 {{\em Phys. Lett. Rev.\/} {\bf #1}, #2}
\def \PASJ #1 #2 {{\em Publ. Astron. Soc. Japan\/} {\bf #1}, #2}
\def \PASP #1 #2 {{\em Publ. Astr. Soc. Pacific\/} {\bf #1}, #2}
\def \NAT #1 #2 {{\em Nature\/} {\bf #1}, #2}
\def \SAIT #1 #2 {{\em Mem.\ Soc.\ Astron.\ It.\/} {\bf #1}, #2}
\def \MESS #1 #2 {{\em The Messenger\/} {\bf #1}, #2}
\def \ASTRNACH #1 #2 {{\em Astron. Nach.\/} {\bf #1}, #2}

\begin{opening}

\title{{\it Beppo}SAX ToO Observations on BLAZARS}
\author{G. Tagliaferri$^{1}$, G. Ghisellini$^1$, M. Ravasio$^{1,2}$}
\institute{$^1$Osservatorio Astronomico di Brera, Via Bianchi 46,
I-23807 Merate, Italy\\
$^2$Universit\`a degli Studi di Milano Bicocca, Piazza della Scienza 3,
I-20126 Milano, Italy}
\date{} 
\end{opening}

\begin{document}

\oddpagefooter{}{}{} 
\evenpagefooter{}{}{} 
\medskip  

\begin{abstract} 
We summarize the results of {\it Beppo}SAX ToO observation of blazars
that were known to be in a high state from observations carried out in the
optical or X-ray or TeV bands. In some of the sources observed, two spectral 
components were detected, which are interpreted as synchrotron and inverse 
Compton emission, respectively. Fast variability was detected in three sources 
(ON\,231, BL\,Lac and S5\,0716+714), but always only for the synchrotron 
component. Most of the triggers are from optical observations, consequently
most of the sources observed are LBL or intermediate objects. They were 
in a high state in the X-ray band, but not in an exceptionally high state.
No strong shift in the synchrotron peak frequency are reported. This is in 
line with the findings that the synchrotron peak frequency is more variable
for HBL objects, i.e. sources that have this peak at higher energies.
\end{abstract}

\medskip

\section{Introduction}

Determining the continuum production mechanism is critical for understanding 
the central engine in AGN, a fundamental goal in extragalactic astrophysics. 
The continuum emission of the Blazar class of Active Galactic Nuclei (AGN) 
is dominated by non-thermal radiation from the radio to the X ray, up to the
MeV, GeV and in some cases TeV energy bands. This emission is often
rapidly variable at all frequencies and, in general, it has been
observed that the amplitude of flux density variations increases and 
the time scales decrease as a function of frequency from the radio to 
the X-ray. A natural explanation is that Blazars are dominated by relativistic 
jets at small angles to the line of sight (Blandford \& Rees 1978; Urry \& Padovani 
1995).

The variability behaviour of a blazar in a given band
depends also from its Spectral Energy Distribution (SED).
It is well known that the Blazar SED is double-peaked (in a $\nu \ vs \ \nu f_{\nu}$
representation). This is interpreted as due to non thermal synchrotron self-compton
(SSC) emission, with the first component due to synchrotron radiation and
peaking at IR to X-ray frequencies and the second one due to inverse Compton 
scattering and peaking in the GeV to TeV band (e.g. Fossati et al. 1998). 
The location of the synchrotron peak is used to define different classes of Blazars:
HBL (High frequency peak Blazars) and LBL (Low frequency peak Blazars) (Giommi \& 
Padovani 1994). In the most studied bands, i.e. radio, optical and X-ray, one expects 
that different sources have different variability behaviour, depending from where 
the synchrotron peak is located.
Normally one expect that the variability is more enhanced after the synchrotron 
peak, towards the end part of the synchrotron emission, where the cooling time 
of the electron is shorter. Longer time scale are expected, and observed, 
in the radio and far-IR bands. Correspondingly, in the X-ray band we expect fast 
variability for sources that hare dominated by the synchrotron emission (HBL), 
while for the sources whose X-ray emission is due to the inverse Compton mechanism 
we do not expect frequent and rapid variability.
Actually, the same source can have the synchrotron peak located at different
frequencies, e.g. in  the presence of flare like events, as the one seen 
in Mkn\,501. This can be explained as due to the injection
of fresh electron in the jet (Pian et al. 1998). 

Since Blazars emit over the entire electromagnetic spectrum, a key for 
understanding blazar variability is the acquisition of several wide
band spectra in different luminosity states during major flaring episodes.
Coupling spectral and temporal information greatly constrains the jet
physics, since different models predict different variability as a
function of wavelength. Before the {\it Beppo}SAX advent, important 
progress in this respect has been achieved for some of the brightest 
and most studied blazars, as PKS2155-304 (Urry et al. 1997), BL\,Lac 
(Bloom et al. 1997), 3C\,279 (Wehrle et al. 1998).
However, thanks to its good energy resolution and sensitivity over an
unprecedented large X-ray energy band, from 0.01 up to 200 keV, {\it Beppo}SAX
immediately provided unique results in the multiwavelength study of Blazar
(e.g. PKS\,2155-304 Giommi et al. 1998, Chiappetti et al. 1999; Mkn\,501 
Pian et al. 1998).
Having in mind all these results we successfully used the {\it Beppo}SAX 
satellite to perform Target of Opportunity (ToO) observations of blazars,
that were known to be in a high state from observations carried out either in 
the optical or X-ray or TeV bands.

\begin{figure}[h,b]
\begin{tabular}{cc}

\epsfysize=8cm 
\hspace{-1.5cm}
\epsfbox{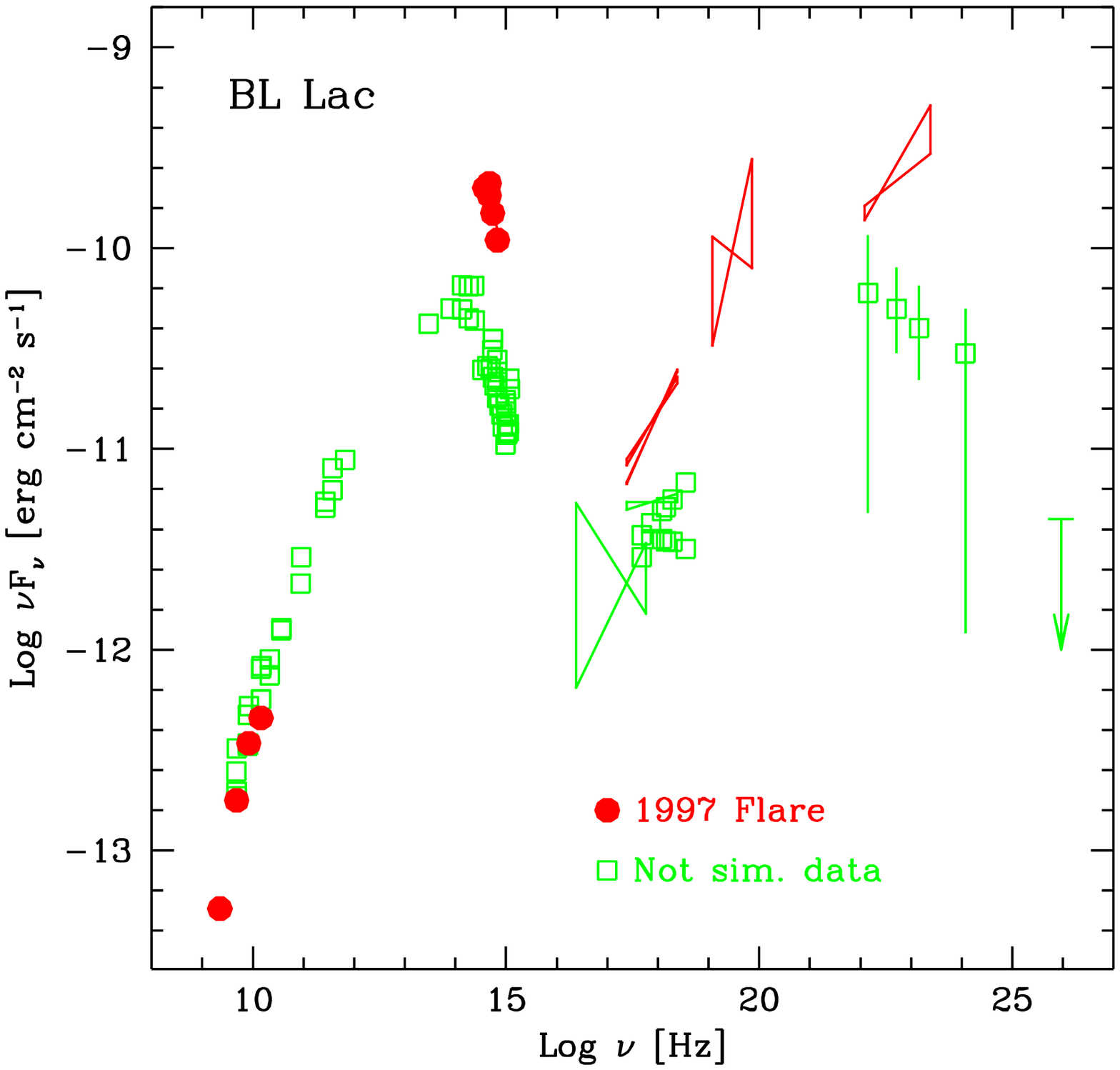}

&

\epsfysize=8cm
\hspace{-1.cm} 
\epsfbox{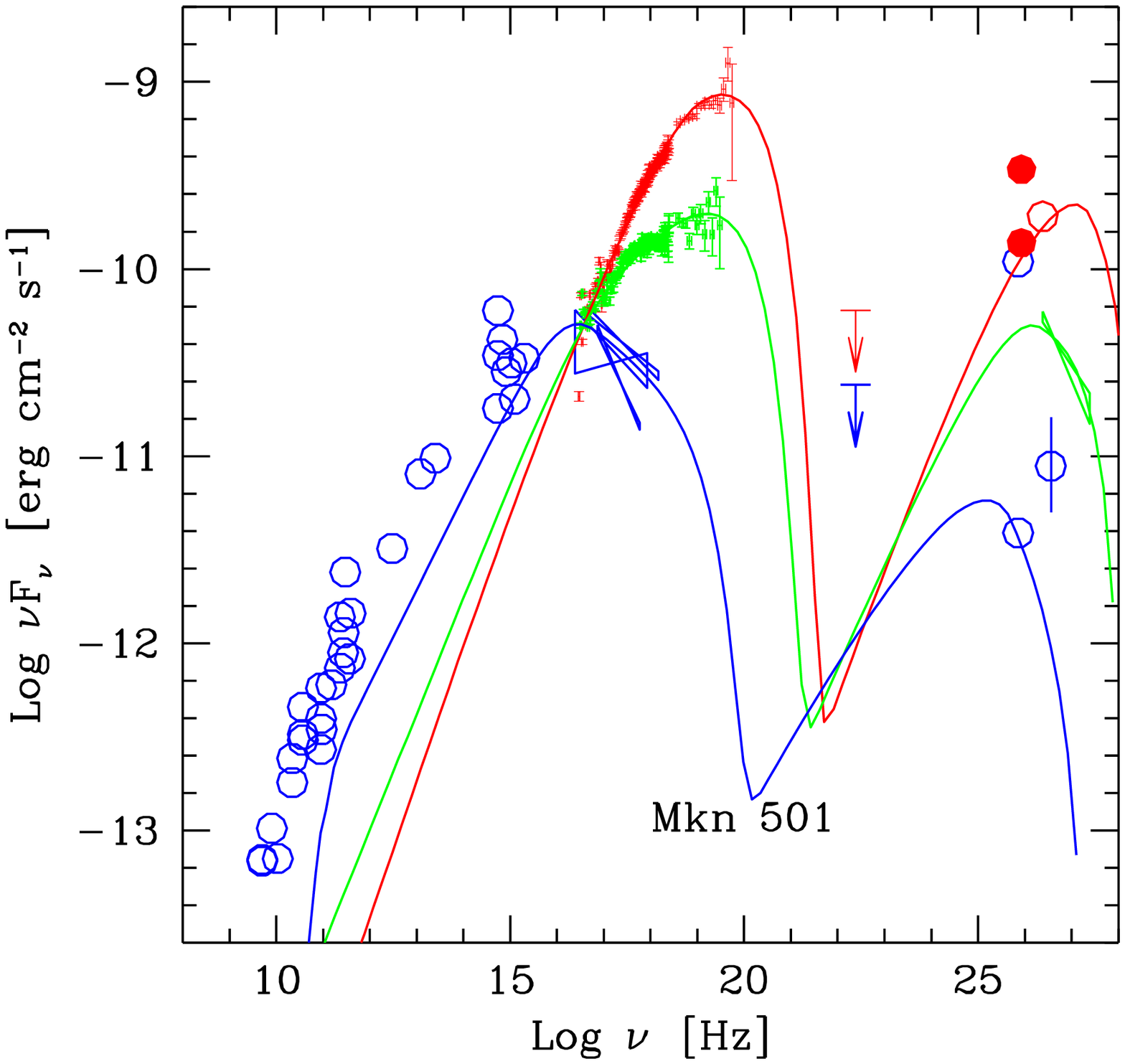}
\end{tabular}
\caption[h]{{\bf Left:} The SED of BL Lac during the flare in the 
summer 1997 (filled symbols, see Bloom et al. 1997) compared to its
`quiescent' level, constructed collecting non--simultaneous data found
in literature. \\
{\bf Right:} Two {\it Beppo}SAX observations of Mkn\,501 while it was in a
high state of activity, compared to previous observations. Note the extremely
large shift of the synchrotron peak toward higher energies.}

\end{figure}

This ToO program was motivated in particular by two spectacular cases. 
The first is the 1997 multiwavelength flare of BL Lac, that we used 
as the paradigmatic case for the optical triggering. During this flare
a number of ground based telescope as well as satellites (ISO, XTE, ASCA
and EGRET) were promptly pointed to BL Lac, triggered by the optical 
observations (IUAC 6693, 6700) of a brightening of over 1 mag.
Data taken within the flaring period are reported in Fig. 1 (left panel,
filled symbols), where they are compared to previous data.
It is evident the increase of the flux at all wavelengths, especially in 
the X--ray band and in the $\gamma$--ray EGRET band, testifying the
large increase of the bolometric luminosity.
Particularly interesting is the behaviour in the X and $\gamma$--ray 
range, where also large spectral variations are evident.
This challenges any model: for instance, in the synchrotron 
self--Compton scenario, an increasing number of emitting electrons 
leads to a linear increase of the optical (synchrotron) flux and to a 
quadratic increase of the X and $\gamma$--ray (Compton) flux, as
observed, but this model does not simply account for the 
flattening of the $\gamma$--ray spectrum and the corresponding shift
of the peak of the Compton component.
The second case was provided by the {\it Beppo}SAX
observations of Mkn\,501.
Quiescent for all 1996, as witnessed by the All Sky Monitor onboard RossiXTE,
at the beginning of 1997 Mkn 501 entered in an extremely high activity phase.
Continuous flaring activity was detected in the TeV band, with flux
levels reaching 4--8 times the level of the Crab.
BeppoSAX observations were scheduled during one of these
flares, leading to the discovery of an unprecedented X--ray emission
for this object (Pian et al. 1998), with a synchrotron spectrum
peaking at or above 100 keV. Compared to previous observations, the peak 
shifted by more than two decades in frequency (see Fig. 1).
This was used as the best case for a X-ray or TeV trigger.

As part of our ToO program we observed 7 different blazars, some of them
more than one time, over a period of 3.5 years. The journal of these
observations are given in Table 1, where we report the source name, the
observation date, the exposure time and the trigger criteria that started
the observation (optical or X-ray, unfortunately we did not have a TeV
trigger). We also report other two ToO observation of Blazars that were 
carried out by {\it Beppo}SAX, but that were not part of our program.
They are Mkn\,421 (Malizia et al. 2000) and OJ\,287 (Massaro et al. 2002).
We will now give the results of some of these observations in more details.

\begin{table}[h,t]
\caption{Journal of the {\it Beppo}SAX Blazars ToO observations}
\begin{center}
\begin{tabular}{|l|c|c|l|}
\hline

Source Name   & Observ. Date & Exposure & Trigger \\
\hline

ON\,231       & 11 May 1998  & 25\,ks   & optical \\
              & 11 Jun 1998  & 32\,ks   &  \\
PKS\,2005-489 & 01 Nov 1998  & 52\,ks   & X-ray \\
BL\,Lac       & 05 Jun 1999  & 54\,ks   & optical+X-ray \\
              & 05 Dec 1999  & 54\,ks   &  \\
OQ\,530       & 03 Mar 2000  & 26\,ks   & optical \\
              & 26 Mar 2000  & 23\,ks   &  \\
S5\,0716+714  & 30 Oct 2000  & 43\,ks   & optical \\
MS\,14588+2249 & 19 Feb 2001 & 48\,ks   & optical \\
1ES\,1959+65  & 23 Sep 2001  & ~7\,ks   & optical \\
              & 28 Sep 2001  & 48\,ks   &  \\
\hline
Mkn\,421      & 22 Jun 1998  & 32\,ks   & X-ray \\
OJ\,287       & 20 Nov 2001  & 40\,ks   & Optical \\

\hline

\end{tabular}
\end{center} 
\end{table}

\section{The Observations}

{\bf ON\,231:} this BL\,Lac object ($z=0.102$), which had been observed
in the X--ray band by
{\it Einstein} IPC in June 1980 with a 1 keV flux of
$1 \mu$Jy (Worrall \& Wilkes 1990) and by ROSAT PSPC in 
June 1991 with a 1 keV flux of $0.4 \mu$Jy and energy spectral
index $\alpha = 1.2$ (Lamer et al. 1996, Comastri et al. 1997),
had an exceptional optical outburst in April--May 1998,
reaching the most luminous state ever recorded, about 40  mJy
in the R band. The optical broad band spectrum was strongly variable.
In particular, it was very flat at the maximum with a broad band 
energy spectral index of 0.52, while before the flare it was found
to be 1.4; the peak 
frequency moved from near IR to beyond the B band. During
the flare a sudden and large increase of the linear polarisation,
from about 3\% to 10\%, was also observed and it remained high at
least to the end of May (Massaro et al. 1999).
Following the optical flare, we triggered our X--ray observation and
ON\,231 was observed by {\it Beppo}SAX in May, with a second pointing
performed a month later, in June. We measured the X--ray spectrum from 
0.1 up to 100 keV. In both occasions the spectrum had a concave shape, 
with a break detected at about 4 and 2.5 keV, respectively.
In Fig.\,2 left panel, we show the SED of ON 231, including our 
simultaneous X--ray and optical data of May, 1998. The SED clearly 
shows that we have detected a concave spectrum in the X--ray band.
We interpret the steeper component at energies below the break as 
due to synchrotron emission and the extremely flat component 
at energies above the break as due to inverse Compton emission. 
This is one of the best example in which both the synchrotron
and the Inverse Compton component are detected simultaneously
and with the same instruments in the X--ray spectrum of a blazar.
In this occasion simultaneous optical observations were also performed.
Unfortunately the source was already close to the sun and it was impossible
to monitor it long enough to search for correlated variability at 
optical and X--ray frequencies.
As shown in Fig.\,3, left panel, during the May observation we
detected a fast variability event with the flux below 4 keV increasing by 
about a factor of three in 5 hours. Above 4 keV no variability was detected.
The X--ray spectra extracted during the flare has (from the third to the 
ninth points of the X--ray light curve shown in Fig.\,3) the first spectral 
index steeper than outside the flare ($\Gamma_1 = 2.7 \pm 0.06$ vs $2.4 \pm 0.15$). 
The break seems to move at higher energies (best fit values are 4.4 and 3.5 keV,
respectively). The second spectral index does not change at all. Thus, 
the fast variability that we detected shows that the break moves at 
higher energies when the source flux increases. The lines shown in the SED 
figure are SSC models that we applied trying to explain the different SED
by changing the minimum numbers of parameters.
The variability predicted by the model can account for the observed 
variability in the soft X--ray band and for the much less variable 
hard X--ray flux, even if the bolometric luminosity does not change
between the two {\it Beppo}SAX observations (note that only the May 1998
observation is reported in Fig. 2 together with a quiescent status
from literature data, clearly there is a change in bolometric 
luminosity between the two SED reported in figure).
This can be achieved by changing (even by a small amount) the slope of
the injected electron distribution, without changing the total injected
power. This will change the synchrotron spectrum above the synchrotron peak,
but not the flux below, nor the self Compton flux below the Compton peak, 
produced by low energy electrons scattering low frequency synchrotron photons
(for a comprehensive analysis of this observation see Tagliaferri et al. 2000).

\begin{figure}[h,t]
\begin{tabular}{cc}

\epsfysize=8cm 
\hspace{-1.5cm}

\epsfbox{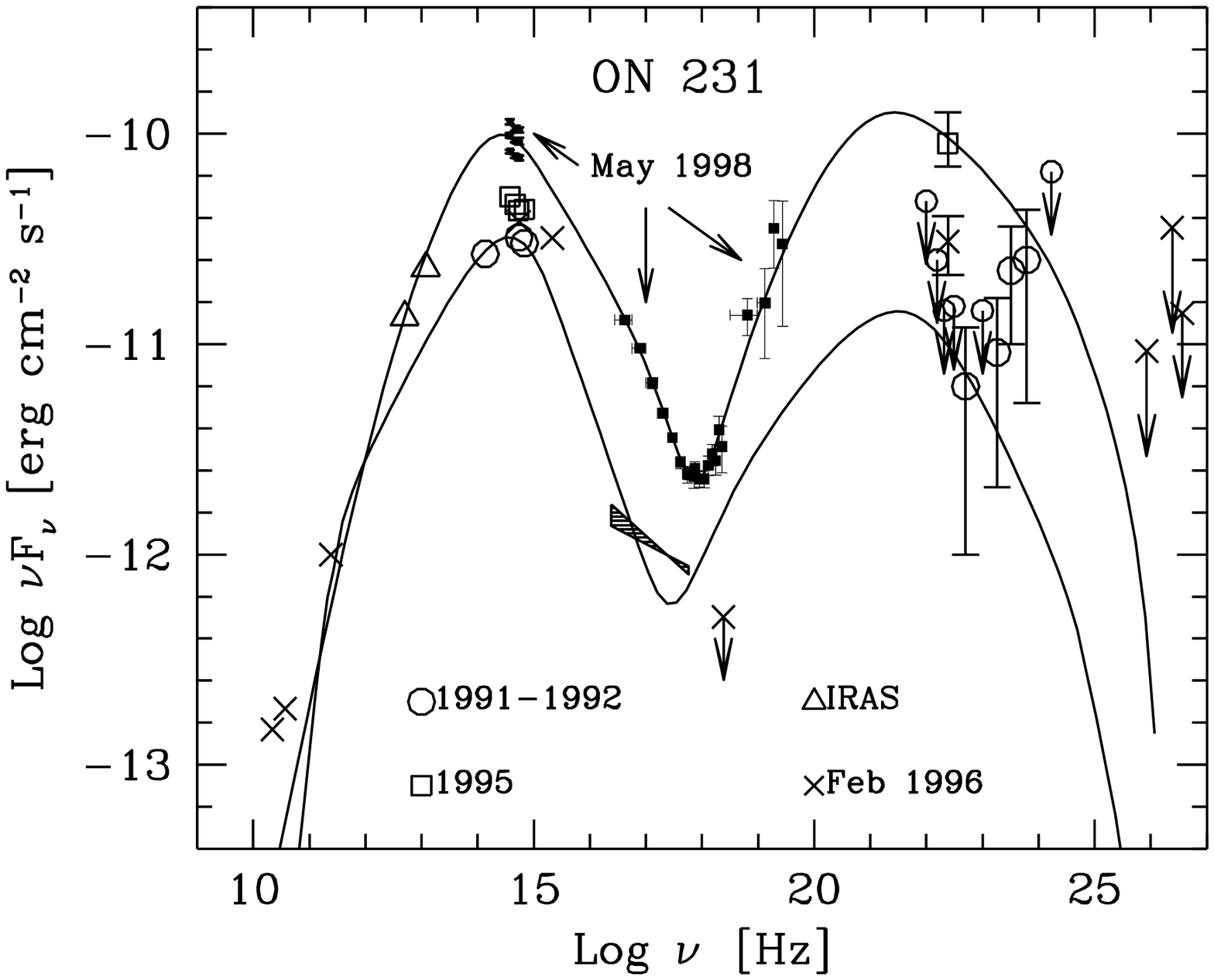}
&

\epsfysize=8cm
\hspace{-1.2cm} 

\epsfbox{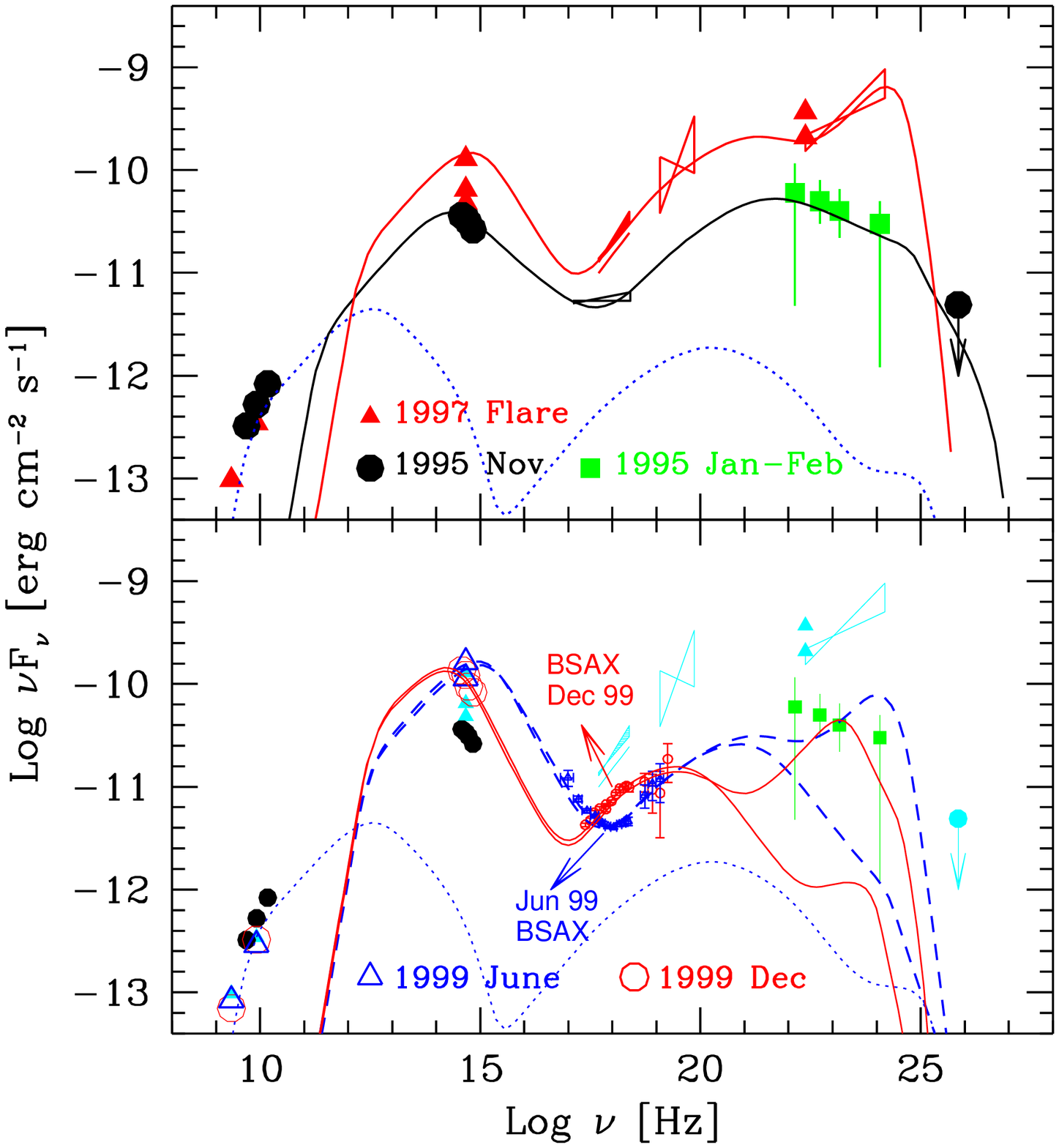}

\end{tabular}

\caption[h]{{\bf Left:} The SED of ON\,231 during the {\it Beppo}SAX
May 1998 observation compared to its `quiescent' level, constructed 
collecting non--simultaneous data found in literature. For more details
about the various data points see Tagliaferri et al. (2000).\\
{\bf Right:} Four simultaneous SEDs of BL\,Lac together with our
best fit models. 
Top panel: the 1997 data modeled with a synchrotron inverse 
Compton model where the emission line photons are important for the 
formation of the very high energy spectrum. 
Bottom panel: triangles and circles correspond to the SEDs of June and December 
1999, as labeled. 
For each SEDs, we show the models corresponding to important or negligible
emission line radiation for the formation of the $\gamma$--ray spectrum.
}
\label{seds}
\end{figure}

{\bf BL\,Lac:} BL Lacertae, being the prototype of the BL \,Lac class, is one of 
the best--studied objects. In the X-ray it has been observed by many satellites
revealing a complex X--ray spectral variability.
In July 1997 following an optical outburst, it was observed
with EGRET, {\it Rossi}XTE and ASCA. EGRET found that the flux level
above 100 MeV was 3.5 times higher than that observed in 1995
(Bloom et al. 1997). {\it Rossi}XTE found
a harder spectrum with a photon index in the range 1.4-1.6 over
a time span of 7 days (Madejski et al. 1999).
A fit to simultaneous ASCA and {\it Rossi}XTE data shows the existence 
of a very steep and varying soft component below 1 keV, photon index 
in the range 3-5, in addition to the hard power law component with
a  photon index of 1.2-1.4. Two rapid flares with time scales of
2-3 hours were detected by ASCA but only in the soft part of the 
spectrum (Tanihata et al. 2000). Finally, in November 1997, BL\,Lac
was observed with {\it Beppo}SAX that detected an energy index of
$0.9 \pm 0.1$ (Padovani et al. 2001).
The two {\it Beppo}SAX ToO observations, performed respectively in 
June and December 1999, were triggered when the source was in very 
high optical states (R $\sim 12.5$), but when the source was actually 
observed by {\it Beppo}SAX the optical flux
was lower (R = 13.4 - 13.6). In both observations the source was
clearly detected up to 100 keV giving us the possibility to study
BL\,Lac over an unprecedentedly large spectral range (0.3-100 keV)
with simultaneous data. BL\,Lac showed quite different spectra: 
in June it was concave with a very hard component above 5-6 keV
($\alpha_1 \sim 1.6$; $\alpha_2 \sim 0.15$); in December
it was well fitted by a single power law ($\alpha \sim 0.6$).
Also in the optical band BL\,Lac showed two different spectral shapes.
Even if optical fluxes are almost similar, the spectrum in June is
harder than in December ($\alpha_{opt} = 1.35$ and 2.02, respectively,
assuming an A$_v=1.09$). 
The synchrotron spectrum of December seems to be shifted 
toward lower energies than that of June: its optical spectrum is softer
and {\it Beppo}SAX detected just the inverse Compton component. 
The observed Compton spectra are quite different, being much harder in
June than in December (see above), and even more so with respect to the
previous {\it Beppo}SAX observation ($\alpha_C= 0.9$, Padovani et al. 2001).
In a simple picture one would have expected that the presence of a break 
in the X-ray spectrum, and hence of a synchrotron component, would have
implied a higher source flux. But this is not the case.
Although the flux below 1.5 keV was higher during the observation 
with the spectral break, the 2-10 keV flux was higher in the other two
{\it Beppo}SAX observations.
All this can be explained by the fact that both the synchrotron, as shown
by the high optical state, and the Compton components are varying. In June, 
the synchrotron flux prevails in the soft X-ray band and two components are 
detected in the X-ray spectrum. However, the 2-10 keV band is still 
dominated by the Compton emission, which is lower but unusually hard.
To show the complex behaviour of the spectral emission of BL\,Lac
we plot in Fig.\,2, right panel, four different SEDs corresponding to 
the multi-wavelength campaigns carried out during November 1995,
during July 1997, when the source was in a very high state and 
during our two {\it Beppo}SAX observations. This figure clearly shows 
the high degree of variability and complexity of BL\,Lac's SED. 
For the details about the SSC model that we applied and that are
shown in the figure see Ravasio et al. (2002).

\begin{figure}[h,t]
\begin{tabular}{cc}

\epsfysize=8cm 
\hspace{-1.5cm}

\epsfbox{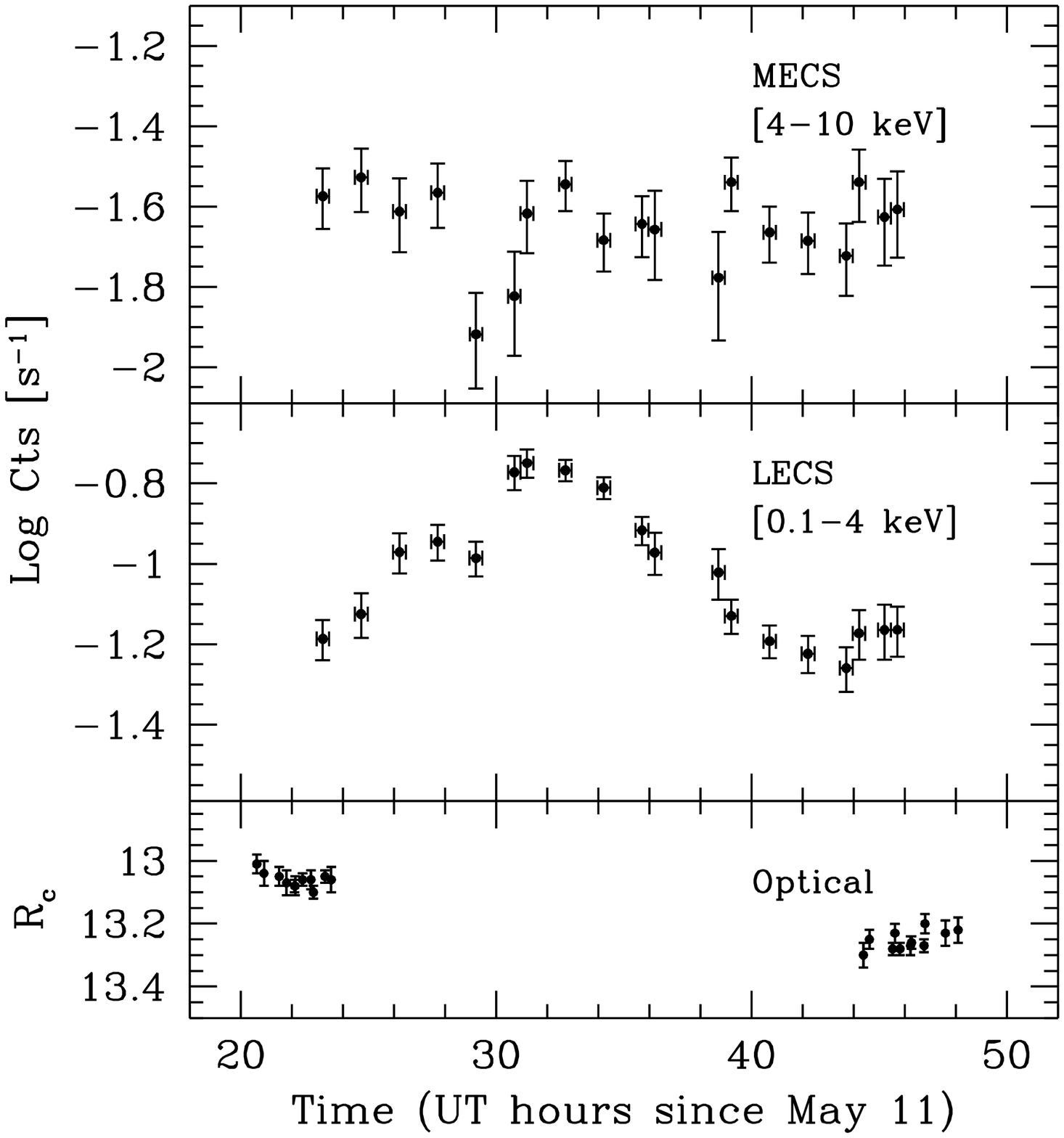}

&

\epsfysize=8cm
\hspace{-1.2cm} 

\epsfbox{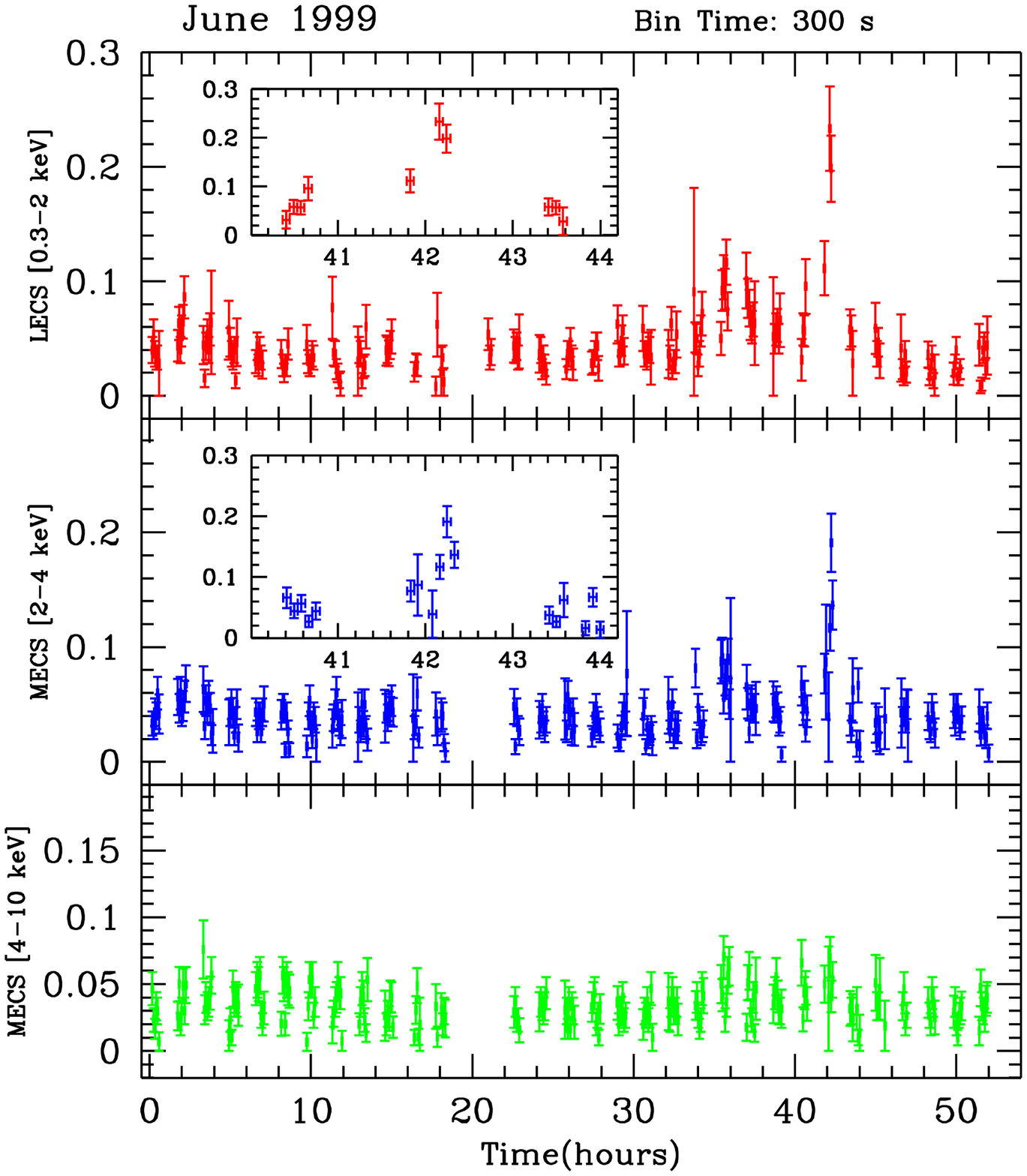}

\end{tabular}
\caption[h]{
{\bf Left:} 4-10 and 0.1-4 keV light curves of ON231 during the May 1998
observation. The two figures have the same scale, note the different amount 
of variability. The optical light curve in the R band is also reported.
{\bf Right:}
LECS 0.3-2 keV, MECS 2-4 keV and MECS 4-10 keV light curves of BL\,Lac during
the June 1999 {\it Beppo}SAX observation. Note the clear flare detected at the 
end of the observations by different detectors (top and mid panels) and better 
shown in the insert. Note also that above the spectral break, i.e. in the 
Compton component, the flare is clearly not present (bottom panel).
}
\end{figure}

The X-ray light curve of the June observation looks in general quite 
constant. However, it reveals an amazing feature about 42 hours after 
the beginning of the observation. In a very short timescale, about
20 minutes, the [0.3-2 keV] flux doubled: this is one of the fastest 
variability ever measured for any BL\,Lacs. In 2 hours the soft X flux 
increased 4 times and then decreased to former values. 
This is not frequency--independent: as can be seen in Fig.\,3, right
panel, variations are detected only in the softer spectral component.
While the synchrotron emission is extremely variable, the inverse Compton 
component remains almost constant during the observation. This behaviour
is very similar to that observed for ON\,231.
Such an amazing event allows us to put severe constraints on the dimension
of the X-ray emission region, on the magnetic field and on
the emitting particle energies.
The frequency dependence is easily explained when performing
the spectral analysis, that highlights a spectral  break attributed to the
transition from the more variable synchrotron to a very hard 
inverse Compton spectrum: synchrotron X-ray emitting electrons are more
energetic than Compton ones, so they cool faster. 
During the second 1999 run, {\it Beppo}SAX detected a softer Compton 
component along its whole spectral range, which accounts for the constance
of the light curves. The comparison of 1995 and 1997 BL \,Lac Spectral 
Energy Distributions, extending to $\gamma$-ray energies,  
suggests the presence of different Compton emission mechanisms. 
We have applied a homogeneous, one--zone synchrotron inverse Compton 
model (SSC) to the four SEDs shown in Fig. 5. We assumed that a spherical
source moves with a bulk Lorentz factor $\Gamma$ along the jet.
We also know that outside the central black hole and accretion disk there
should be the broad line region zone. Infact emission lines have been detected
in the optical spectra of BL\,Lac, which have some times exceeded the 
canonical threshold of $5 \ \AA$ of equivalent width to define an objects
as a BL\,Lac (Vermeulen et al. 1995). Thus, in our model the seed soft
photon distribution is provided by the synchroton process only if the 
source is located at a distance, along the jet, greater than a critical 
distance $z_{BLR}$ where most of the line emission is produced.
Otherwise, we include the radiation energy density of the emission
lines, as seen in the frame comoving with the source. The applied model 
is aimed at reproducing the spectrum originating in a limited part of
the jet, thought to be responsible of most of the emission. This region
is necessarily compact, since it must account for the fast variability.
In our model we assumed that the dissipation region of the jet
corresponds to the collision of two blobs or shells moving at slightly
different velocities. This is the ``internal shock" scenario for blazars,
as proposed by Ghisellini (1999) and Spada et al. (2001) (see also
Madejski et al. 1999). 
We therefore have that different SEDs correspond to different
collisions, that can occur at different location in the jet. 
Some of them may be located within the size of the BLR, while
others occur outside. This has a quite dramatic effect on the 
spectrum, since within the BLR the energy density of the emission 
line radiation easily exceeds the magnetic and the synchrotron energy
densities. Therefore, collisions occurring at $z<z_{BLR}$ emit 
most of their power in the GeV band, as happened in the 1997 flare.
On the contrary, for collisions beyond $z_{BLR}$ the emission
line radiation is negligible, and the Compton to synchrotron luminosity
ratio is controlled by the ratio between the magnetic field and the 
synchrotron energy densities. This is the most plausible scenario for
the 1995 SED, characterised by the fainter and steeper EGRET spectrum.
For the 1999 June SED, the very short variability timescale suggests a 
very compact region, and hence a location for the inverse Compton region 
within the Broad Line Region, closer to the jet apex. For the 1999 
December observation we do not have tight constraints from variability
to discriminate if regions are within or outside the BLR.
For more details about the model see Ravasio et al. (2002).

{\bf PKS\,2005-489:} this bright BL\,Lac object was observed
following an active X-ray state detected by {\it Rossi}XTE.
This is the only X-ray trigger of our ToO campaign. The source 
was in a very high state with a continuum, detected up to 200 keV,
well fitted by a steepening spectrum due to synchrotron emission only.
We did not detect fast variability during the observation. Our X-ray 
spectrum is the flattest ever observed for this source, with a 
synchrotron peak frequency located between $10^{15}$ and 
$2.5 \times 10^{16}$ Hz, depending on the model assumptions.
Although the source has been observed over a large range of X-ray 
fluxes, the different X-ray spectra, as measured by various satellites, 
are consistent for this object with relatively little changes of the
peak frequency of the synchrotron emission, always located below
$10^{17}$ Hz (for more details see Tagliaferri et al. 2001).

{\bf S5\,0716+714 \& OQ\,530:} these two sources were both observed in 
the year 2000 due to optical triggers. For both of them we detected
a break in the spectrum that again can be interpreted as due to synchrotron 
and inverse Compton emission. In the case of S5\,0716+714 we detected
again fast variability only in the synchrotron part of the spectrum. This
different variability behaviour of the two components was already detected
in a previous {\it Beppo}SAX observation of this source (Giommi et al. 1999).
For both sources our observations have the highest flux, when compared with 
previous X-ray observations, but we have no indication of a strong variability
in the SED of both sources. We have also simultaneous TeV observation
performed with the HEGRA telescope, but both sources are not detected
(Tagliaferri et al. 2002).

{\bf OJ\,287 \& MS\,14588+2249:} these sources were observed in 2001 as
part of two different ToO programs. Also in these two cases the triggers 
were from optical observations. However, both sources were detected in
a low status in the X-ray and their spectra are well represented by a single
powe law model. For OJ\,287 the spectrum is flat $\alpha_x = 0.45 \pm 0.08$
and due to Compton emission, while for MS\,14588+2249 it is steeper,
$\alpha_x = 1.71 \pm 0.15$ and due to synchrotron emission
(for more details see Massaro et al. 2002).

{\bf Mkn\,421:} this is one of the most studied BL\,Lac object in the 
X-ray band. Spectacular variability has been detected during long 
{\it Beppo}SAX observations, in one case, in April 1998, simultaneously 
with a TeV flare, detected also in the X-ray band.
These data show that the X-ray and TeV intensities are well correlated
on timescales of hours, implying that the X-ray and TeV photons derive
from the same region and from the same population of relativistic
electrons (Maraschi et al. 1999). The good statistic and large
{\it Beppo}SAX energy band allowed Fossati et al. (2000) to detect
the peak of the synchrotron component shifting to higher energies 
during the rising phase and then decreasing to lower energies during
the decay phase. In June of the same year a ToO observation was 
triggered when Mkn\,421 was detected to be in a high state by one 
of the {\it Beppo}SAX Wide Field Camera (Malizia et al. 2000). 
Also during this ToO observation
the source showed short-term temporal and spectral variability.
The source hardens while brightening and the synchrotron peak
moves to higher energies, in agreement with previous results and 
in particular with the findings of Fossati et al. (2000).
In Fig. 4 we report the 0.1-100 keV $\nu f_{\nu}$ spectra of 
Mkn\,421 recorded during the ToO and two previous {\it Beppo}SAX
observations when the source was weaker (from Malizia et al. 2000).
This clearly shows that the synchrotron peak moved to higher 
frequencies when the flux increased.

\begin{figure}[h,t]
\begin{tabular}{cc}

\epsfysize=6cm 
\hspace{-0.5cm}

\epsfbox{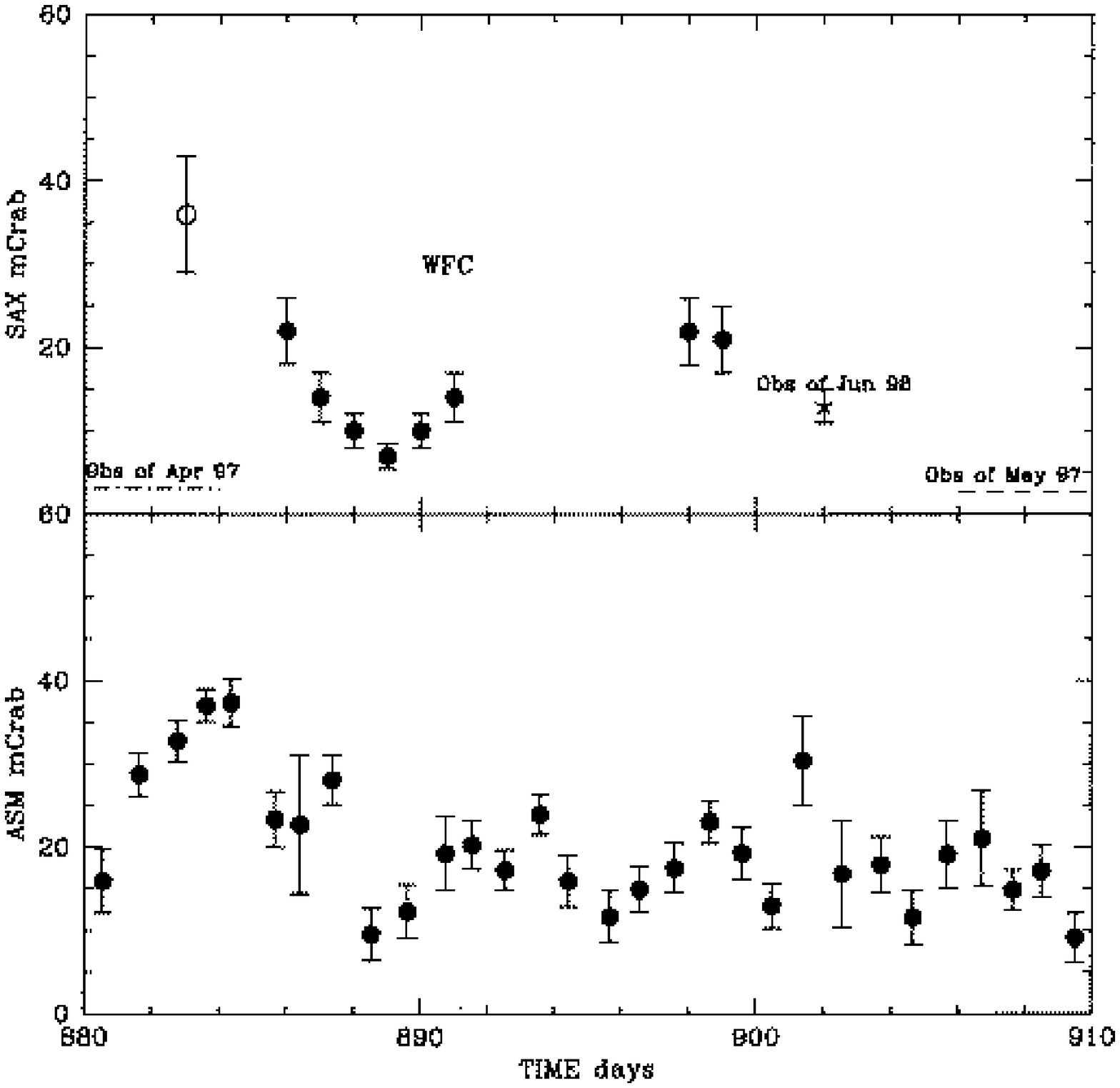}

&

\epsfysize=6cm

\epsfbox{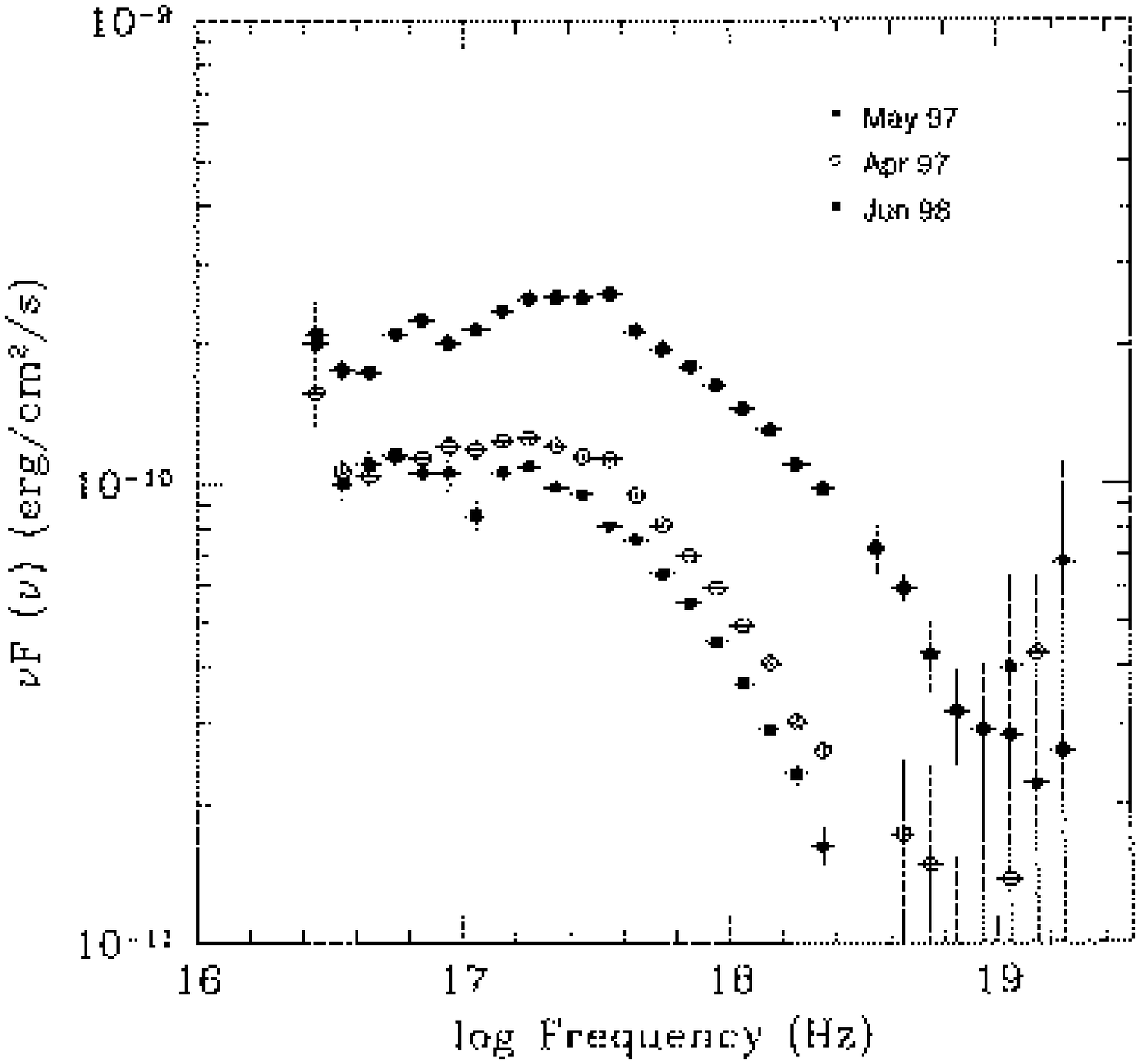}

\end{tabular}
\caption[h]{
{\bf Left:} {\it Beppo}SAX WFC light curve (top panel) compared
with RXTE ASM one-day average (bottom panel) during the {\it Beppo}SAX
observation of 1998 June. In the top panel, the dashed lines identify
the flux levels recorded during the two {\it Beppo}SAX exposure in 1997.
The star indicates the flux seen during the June 1998 observation.
{\bf Right:} 0.01-100 keV spectra of Mkn\,421 during three {\it Beppo}SAX
observations. Note how the peak of the emission moves to higher energies 
when the flux increases. Both figures are from Malizia et al. (2000).}

\end{figure}

\section{Conclusion}

We presented some of the most important results
that were obtained with {\it Beppo}SAX ToO observations of Blazars.
They immediately show the importance of observing in the X-ray band
Blazars that are known to be in a high state. With {\it Beppo}SAX this
is even more true: thanks to its large energy range, it has been possible
to simultaneously detect both the synchrotron and the Compton components.
We detected fast variability events that allowed us to put constraints 
on the size of the emitting region and to infer the properties of the jets
responsible for the Blazars' emission.

But we had also limits in this projects. For instance, it can be seen 
from Table 1 that the Blazars ToO program has been dominated by optical 
triggers. Thus, we are probably biased towards sources that have
an higher optical variability. These should be the blazars that have
the synchrotron peak in the IR-optical band, i.e. sources with the peak
on the left side of the bands in which the blazars are
usually monitored. These are of course essentially LBL
or intermediate blazars. Sources that have the synchrotron peak in the 
UV--X-ray band are not expected to show strong optical variability.
They are probably more easily detected in a high state from systematic 
monitoring at higher frequencies. Of course this is much more difficult 
than in the optical and this explain while in Table 1 there are only two 
X-ray triggers. This could explain the fact that most
of the sources observed as ToO observations were in a high state in
the X-ray band, but not in an extremely high state. Moreover, we note
that from the optical trigger to the actual X-ray observations normally
there are delays of a few days and this could also explain while we did
not find sources in exceptionally high states.

\begin{figure}[h,t]
\vspace{-2.0cm}

\epsfysize=9cm

\epsfbox{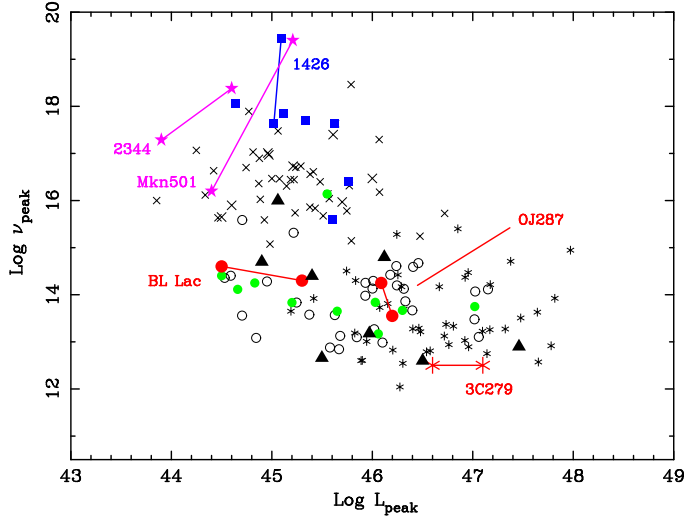}

\caption[h]{Peak frequency vs. luminosity at the peak frequency
for a sample of Blazars. Note how the high peak objects seem to show
higher variability of the synchrotron peak frequency, while lower peaked
sources show a rather steady $\nu_{peak}$ among different luminosity
states (from Costamante et al. 2000). }

\end{figure}

These observations shows also that the synchrotron peak frequency
does not seem to vary a lot in the LBL or intermediate objects.
On the contrary, strong shifts have been observed for Mkn\,501 and 
Mkn\,421 that are HBL. This is in line with the findings of 
Costamante et al (2001), that studied with {\it Beppo}SAX blazars
with extreme synchrotron peak frequencies ($\nu_{peak} > 1$ keV).
They found that these sources seems to be characterised by larger
$\nu_{peak}$ variability, compared with lower $\nu_{peak}$ objects
(see Fig. 5).

We detected strong spectral variability, founding in the same source
either two or only one component. We also found fast 
variability, but only in the synchrotron component of sources showing 
both the synchrotron and the Compton components. Fast variability is
present also in the optical band, but it is less pronounced and there
is no one to one correspondance. All this can be intepreted with the
presence of a steady Compton component, and the erratic variability of
the synchrotron tail emission, coming in and out of the soft X-ray band.
The Compton emission we see in the X-ray band is well below the Compton peak 
and it is produced by low energy electrons scattering low frequency synchrotron
photons. The variability seen in the synchrotron part can be obtained
by changing the slope of the injected electron distribution, without
affecting the total injected power. Time to time there are variability
also in the Compton component and this imply a strong modification
of the overall blazar SED, as in the case of the 1997 BL\,Lac flare.
All these behaviours can be reproduced by the presence of relativistic jets 
dominated by shock events produced by colliding shells (e.g. Spada et al. 2001).

In any case these observations show the importance of observing Blazars 
over such a large X-ary energy band while they are in a high state. 
It will be crucial to perform these ToO observations also in the 
future and in particular for Blazars that are detected in a high state
in the X-ray or TeV bands, now that new and more sensitive TeV telescope
will be operational. In the forseable future these ToO observations,
as the one carried out with {\it Beppo}SAX, will be possible probably 
either with the combination of simultaneous observations from Integral 
and other soft X-ray satellites (such as XMM-Newton or Chandra), or
with the Swift satellite. Swift will be launched at the end of 2003.
Swift is dedicated to the study of Gamma Ray Bursts, but it
should also observe other interesting sources, whose emission
goes from the soft to the hard X rays, and Blazars are obvious
good candidates.

\acknowledgements
This research was financially supported by the Italian Space 
Agency and MIURST. We thank the {\it Beppo}SAX Science Data
Center (SDC) for their support to this program.


\end{document}